\documentclass[aps,prl,twocolumn,letterpaper,longbibliography]{revtex4-2}
\usepackage{graphicx}
\usepackage{bm}
\usepackage{amssymb}
\usepackage{color}
\usepackage{amsmath}
\usepackage{ulem}
\usepackage{bbold}
\usepackage[mathlines]{lineno}
\usepackage[colorlinks=True,citecolor=blue,linkcolor=blue,allcolors=blue]{hyperref}
\usepackage{enumitem}
\usepackage[utf8]{inputenc}

\begin{document}
\title{Tunable two-dimensional laser arrays with zero-phase locking}

\author{Wei Xin Teo}
\thanks{These authors contribute equally.}
\affiliation{Department of Physics, National University of Singapore, Singapore 117542, Singapore}
\author{Weiwei Zhu}
\thanks{These authors contribute equally.}
\email{phyzhuw@gmail.com}
\affiliation{Department of Physics, National University of Singapore, Singapore 117542, Singapore}
\author{Jiangbin Gong}
\email{phygj@nus.edu.sg}
\affiliation{Department of Physics, National University of Singapore, Singapore 117542, Singapore}
\
\date{\today}

\begin{abstract}
Two-dimensional (2D) laser arrays are shown to be achievable at a large scale by exploiting the interplay of higher-order topological insulator (HOTI) physics and the so-called non-Hermitian skin effect (NHSE). The higher-order topology allows for the amplification and hence lasing of a single-mode protected by a band gap; whereas the NHSE, widely known to accumulate population in a biased direction in non-Hermitian systems, is introduced to compete with the topological localization of corner modes. By tuning the system parameters appropriately and pumping at one site only, a single topologically protected lasing mode delocalized across over two dimensions emerges, with its power widely tunable by adjusting the pump strength.  Computational studies clearly indicate that the lasing mode thus engineered is stable, and the phase difference between nearest lasing sites is locked at zero, even after the disorder is accounted for. The total power of the lasing mode forming a 2D topological laser array is proportional to the area of the 2D lattice accommodating a HOTI phase. Based on existing experiments, we further propose to use coupled optical ring resonators as a promising platform to realize large-scale 2D laser arrays.
\end{abstract}

\maketitle

{\it Introduction.}---Topological lasing is an appealing concept that advocates laser operation in topological systems with optical gain and loss, offering  robustness against disorder due to topological protection \cite{Pilozzi2016,St-Jean2017,longhi2018non,longhi2018invited,Malzard_2018,Malzard:18,zapletal2020long,amelio2020theory,zhao2018topological,PhysRevLett.120.113901,ota2018topological,suchomel2018platform,shao2020high,gao2020dirac,ezawa2021nonlinear}. Lasing with topological edge modes was first proposed and experimentally realized in a one-dimensional (1D) Su-Schrieffer–Heeger (SSH) model \cite{Pilozzi2016,St-Jean2017}. Because the topological edge mode is exponentially localized at the end of the 1D SSH chain, the power of such lasing is limited by construction.  Topological lasing in 2D geometries is hence of importance in connection with the possibility of realizing large-scale topological laser arrays with a higher power. Indeed,  recent studies of topological lasing have involved different types of 2D topological lattices,  including those based on Chern insulator, valley topological insulator, or topological crystal insulator \cite{harari2016topological,bahari2017nonreciprocal,doi:10.1126/science.aar4003,doi:10.1126/science.aar4005,longhi2018presence,klembt2018exciton,kartashov2019two,PhysRevResearch.1.033148,gundougdu2020edge,noh2020experimental,zeng2020electrically,Choi2021,yang2020mode,ishida2021large}.  These inspiring attempts have all demonstrated the lasing along certain topological edge modes extended along the edge of a 2D lattice, with the lasing power much enhanced when compared to topological lasing based on zero-dimensional edge modes.
%Furthermore, single-mode lasing, which is also a crucial property for a high-power coherent laser source, can also be achieved \cite{doi:10.1126/%science.aar4003,doi:10.1126/science.aar4005,yang2020mode,ishida2021large,qiao2021higher}.
Nevertheless, in most current 2D topological lasing examples proposed to date, the bulk lattice sites are not directly participating in the lasing.  A genuine 2D array of zero-phase-locked topological lasers with widely tunable output power  is not materialized yet.

In this work, we show how a 2D array of zero-phase-locked topological lasers can be engineered with widely tunable output.  The starting point of our design is
the use of delocalized topological corner modes (TCMs) in the band gap.  At first glance, a delocalized TCM for lasing
sounds self-contradictory, but it is in fact feasible by the interplay of higher-order topological insulator (HOTI) physics \cite{Benalcazar2017,Peterson2018,Serra-Garcia2018,Imhof2018,Mittal2019,Xie2019,Chen2019,Liu2019,Xue2018,Ni2018,Zhang2019,Zhu2021,Zhu2022} and the so-called non-Hermitian skin effect (NHSE) \cite{Yao2018,Yao2018a,Lee2019a,Kawabata2019,Lee2019,Brandenbourger2019,Okuma2020,Zhang2020a,Weidemann2020,Li2020}.   This then results in a significant enhancement of the output power of single-mode lasing because now many sites over a 2D lattice play an equal role in the lasing.
%Our main findings are hence remarkably different from previous studies of 2D topological lasers with TCMs of HOTIs \cite{zhang2020low,han2020lasing,kim2020multipolar,zhong2021theory,zhu2021single,ezawa2021nonlinear}, where the power of topological lasing is still limited due to the topological localization of TCMs.
Compared with the existing proposals~\cite{zhang2020low,han2020lasing,kim2020multipolar,zhong2021theory,zhu2021single,ezawa2021nonlinear,Teimourpour2016,Wong2021}, the single-mode lasing here with single-site pumping is maintained across a wide range of pump strengths and hence a wide range of output power.  Furthermore,  all the lasing sites are locked with zero relative phase and can be easily scaled up across two dimensions, even in the presence of disorder.

\begin{figure}[h]
    %\centering
    \includegraphics[width=0.7\linewidth]{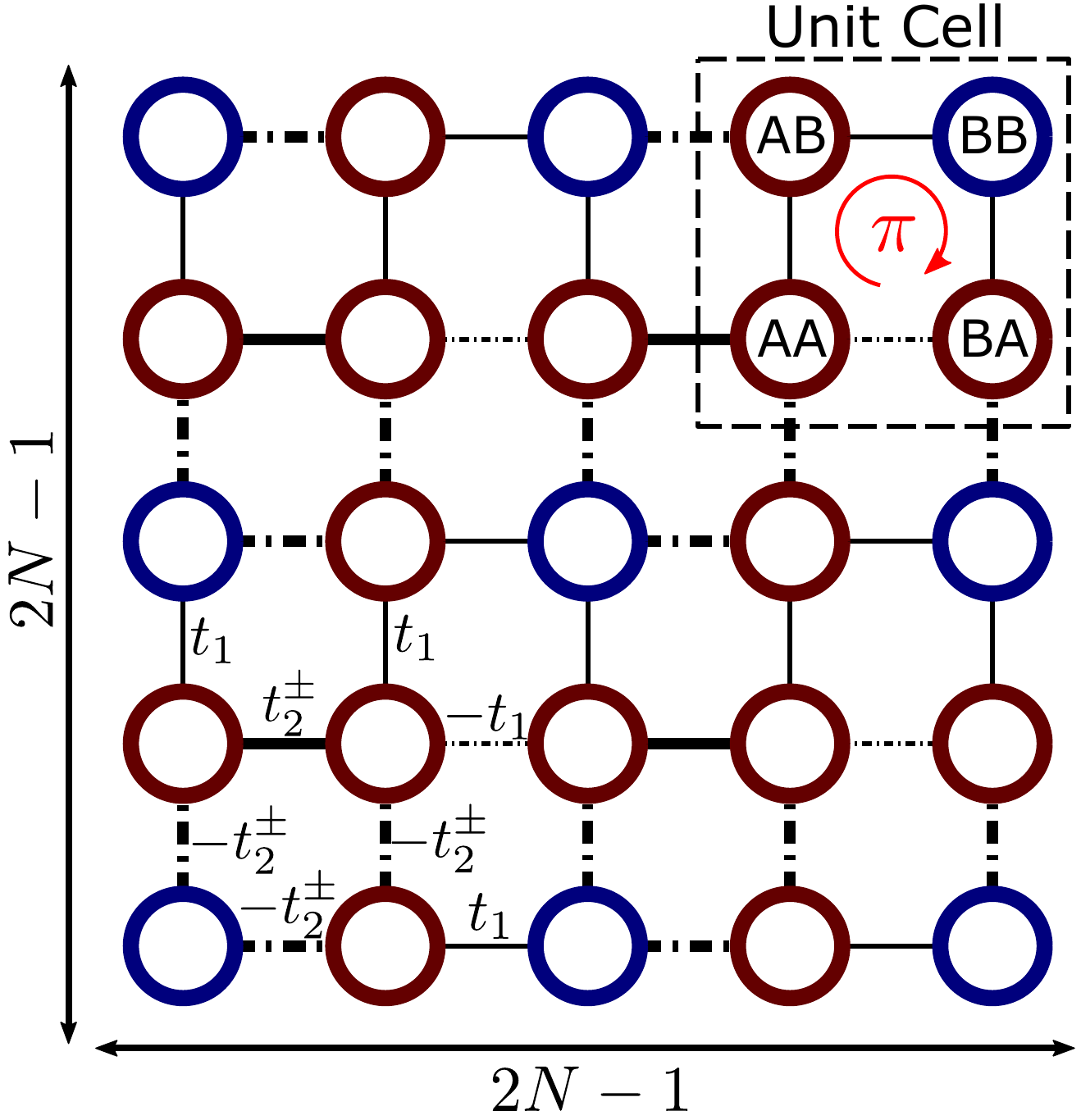}
    \caption{Schematics of a 2D lattice of coupled ring resonators accommodating a TQI phase, made possible by engineered couplings. The unit cell is boxed by the dashed line, which consists of 4 sites (AA,BA,AB,BB). Along each direction, there are $(2N-1)$ lattice sites. The intercellular coupling $t_2^\pm=t_2e^{\pm g}$ is nonreciprocal, with the couplings  $t_2e^{+g}$ pointing towards the left and the bottom. Note also that along the ordinate, the signs of the intracellular coupling and intercellular coupling are alternating.}
    \label{fig:model}
\end{figure}

{\it Model.}---  Several studies have curiously shown that the topological edge mode of a 1D lattice can be made completely delocalized due to a balanced competition between topological localization and the biased population accumulation of NHSE \cite{longhi2018non,gao2020anomalous,zhu2021delocalization,Yan2021}.  This hints the possibility of having completely delocalized TCMs for lasing considerations.  We find that this is indeed the case by investigating a lattice hosting the topological quadrupole insulator (TQI) phase.   To facilitate future experimental implementation, we consider an explicit context, namely, a coupled-ring system that is made of active resonators with the following Hamiltonian \cite{doi:10.1126/science.1258479,doi:10.1126/science.1258480,doi:10.1126/science.aar4003,ezawa2021nonlinear,zhu2021single}:
\begin{equation}
\label{eq1}
    H=H^{\rm hop}+\sum_n c_n^\dag c_n (-i\gamma+P(n)),
\end{equation}
where $c_n$ and $c_n^\dag$ are the annihilation and creation operators, $\gamma$ represents the loss in each resonator, and $P(n)$ describes a site-dependent optical saturation gain via stimulated emission.  $H^{\rm hop}$ here describes the couplings between the resonators.
As one example, here we assume that the resonators are arranged in square lattice arrays. To obtain a TQI phase, the couplings between the resonators are engineered to include both positive couplings and negative couplings, and each four-site plaquette has $\pi$ flux, as shown in Fig.~\ref{fig:model} \cite{Benalcazar2017,zhu2020distinguishing}.  For NHSE to play a role,  we also introduce the asymmetric couplings between the unit cells. Experimental realizations of this design will be elaborated on later.

With these preparations, the above $H^{\rm hop}$, as divided into $x$-direction part and $y$-direction part, is further assumed to be as follows:
\begin{equation}\label{eq2}
 H^{\rm hop}=H^x+H^y,
\end{equation}
with each part constructed from a 1D SSH model with some nonreciprocal intercellular coupling,
\begin{eqnarray}
 H^x &=& (I_{N\times N}\otimes\sigma_3)\otimes H^{\mathrm{SSH}} \label{eq3} \\
 H^y &=& H^{\mathrm{SSH}}\otimes (I_{N\times N}\otimes \sigma_0)  \label{eq3n}
\end{eqnarray}
where $\otimes$ is the Kronecker product, $I_{N\times N}$ and $\sigma_0$ are the identity matrices of dimension $N$ and $2$ respectively; $\sigma_{3}$ in Eq.~(\ref{eq3}) is the third Pauli matrix to reflect the alternating signs of couplings in odd row and even row as shown in Fig.~1; $H^{\mathrm{SSH}}$ in Eq.~(\ref{eq3}) represents SSH-type of coupling along the $x$ direction, whereas $H^{\mathrm{SSH}}$ in Eq.~(\ref{eq3n}) depicts SSH-type of coupling along the $y$ direction, with the 1D $H^{\mathrm{SSH}}$ defined as
\begin{equation}
\label{eq4}
\begin{split}
    H^{\mathrm{SSH}} = & \sum_{i=1}^{N} [t_1 |i\rangle\langle i| \otimes \sigma_1] - \sum_{i=1}^{N-1}  [t_2e^g |i\rangle \langle i+1| \otimes \sigma_+ \\
    & + t_2e^{-g} |i+1\rangle \langle i| \otimes \sigma_-],\\
\end{split}
\end{equation}
where $|i\rangle=c^\dag_i|0\rangle$, $t_1$ denotes the intracellular coupling, $t_2$ denotes the intercellular coupling, $g$ denotes the strength of non-reciprocity, $N$ is the number of unit cells along one side of the square lattice system, and $\sigma_{\pm}=(\sigma_1\pm\sigma_2)/2$ with $\sigma_{1,2}$ are the first two Pauli matrices. As usual, the pseudo-spin here refers to the sublattice degree of freedom in the SSH model.
 The 1D SSH element of our design is motivated by the fact that such a 1D SSH model with nonreciprocal couplings can support a single delocalized topological zero mode in the band gap, provided that the condition $|\ln(t_1/t_2)|=g$ is met ~\cite{zhu2021delocalization}.    Let us use $\{x,y\}$ to denote the coordinate of the lattices, i.e. $\{1,1\}$ denotes the lattice at the left bottom corner.  For systems under the open boundary condition,  we set the number of sites along each direction to be $(2N-1)$ (so the last unit cell is cut on purpose).  For convenience we also use one single lattice site number $n$ to refer to individual lattice sites on the square geometry,  via the following simple relations: $n=(y-1)(2N-1)+x$, $y=[\frac{n-1}{2N-1}]+1$, $x=n\mod{(2N-1)}$, with
$[\ ]$ being the floor function.  With this definition of $n$, the site-dependent optical saturation gain $P(n)$ can be written as follow:
\begin{equation}\label{eq6}
    P(n)=\delta_{n,\bar{n}}(\frac{i\gamma\xi}{1+|\psi_n|^2/\eta}),
\end{equation}
where $\bar{n}=(2N-1)^2$.  This is an important detail of our study. Indeed, this expression indicates that we only apply the gain at the top right corner of the 2D lattice, so that only the top right TCM is stimulated.  Note also that $\gamma\xi$ represents the amplitude of the optical gain, and $\eta$ represents the non-linearity, with the system approaching linear if $\eta\xrightarrow{}\infty$.

\begin{figure}[h]
     \centering
     \includegraphics[width=0.5\textwidth]{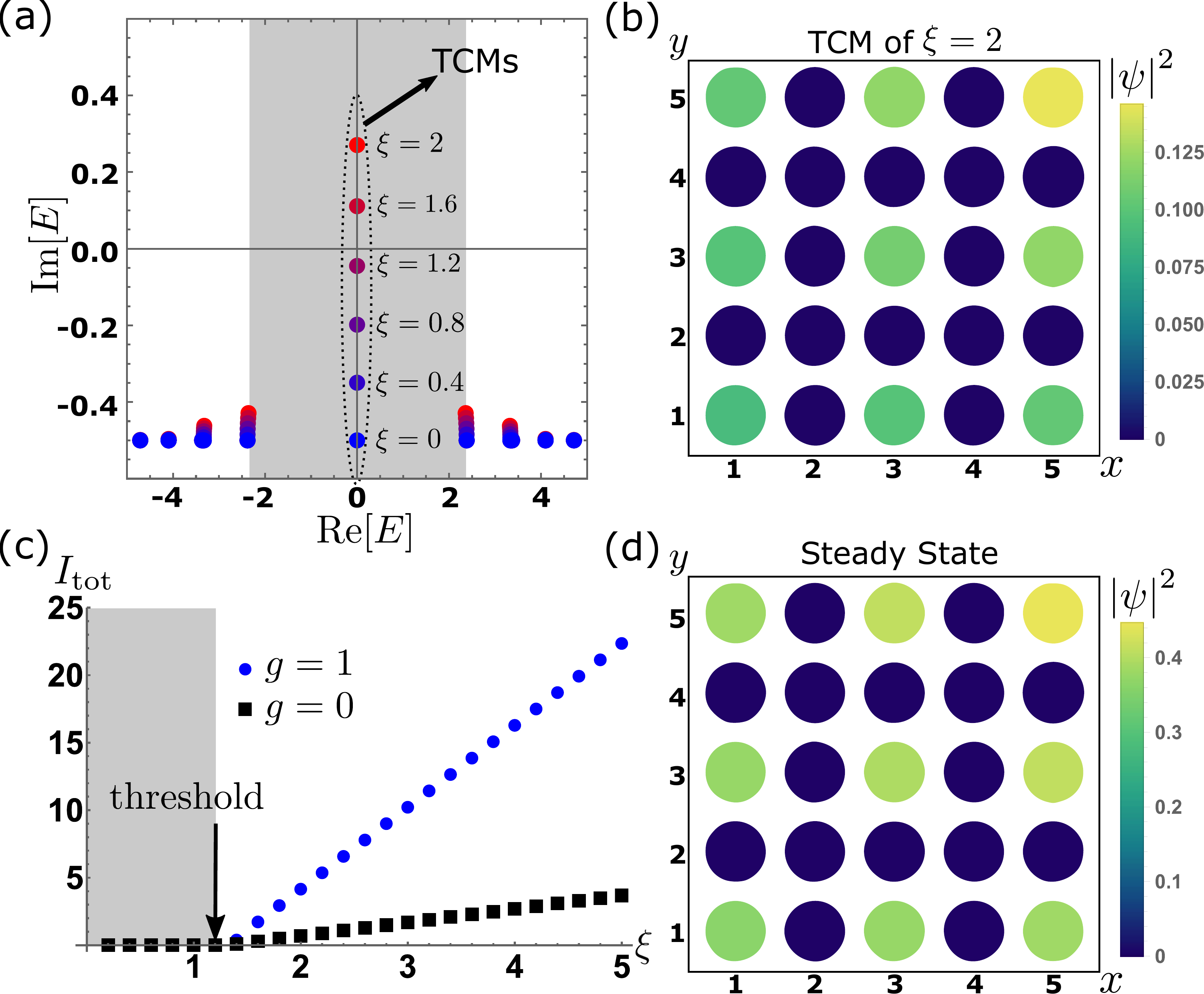}
     \caption[justification=justified]{(a) Spectrum of the designed Hamiltonian shown in Eq.~(\ref{eq1}), with different values of the optical gain amplitude parameter $\xi$.  A TCM with zero real energy is always seen to be located within the band gap. For $\xi$ not sufficiently large, the energy of the TCM has a negative imaginary part, and hence this mode cannot be significantly stimulated. (b) The density profile of the TCM for $\xi=2$. The TCM has significant state densities extended over the ``BB" sites, constituting strong evidence that the in-gap TCM is delocalized across the entire 2D lattice. (c) The total saturated intensity $I_{\rm tot}$ vs $\xi$ for $g=1$ and $g=0$ respectively.  $\xi\approx1.3$ is seen be the threshold value for the TCM to be stimulated. Beyond this threshold, the total intensity with NHSE ($g=1$) increases much faster than the case without NHSE ($g=0$). (d) The density profile of the steady state after a long time evolution with $\xi=2$. Unless specified otherwise, system parameters here are set at $N=3$, $t_1=1$, $t_2=e$, $g=1$, $\gamma=0.5$, and $\eta=1$. }
    \label{fig:delocalized}
\end{figure}

{\it Single topological mode delocalized over 2D}---To achieve single-mode lasing across a 2D geometry, it is important to first engineer a topological lattice with a single TCM.  It is this motivation that calls for our construction in Eq.~(\ref{eq3}). A pioneering study of 2D laser arrays~\cite{qiao2021higher} used two commuting components as well as supersymmetry,  without topological protection or connection with TCMs.
By contrast, in our design $H^{x}$ anti-commutes with $H^{y}$.
It comes from the fact that the 1D SSH model is chiral symmetric, i.e. $\Gamma H^{\mathrm{SSH}} \Gamma^\dag = -H^{\mathrm{SSH}} $ with $\Gamma=I_{N\times N}\otimes\sigma_3$.
With this, it can be proved easily that the first part of the $H^{x}$ and $H^{y}$ anti-commutes with each other, with $\{I_{N\times N}\otimes\sigma_3,H^{\mathrm{SSH}}\}=0$, while the second part commutes with each other, with $[I_{N\times N}\otimes\sigma_0,H^{\mathrm{SSH}}]=0$. Therefore, by combining the two parts, we obtain $\{H^{x},H^{y}\}=0$.
This being the case,
\begin{equation}\label{eq7}
  H^{\rm hop}\cdot H^{\rm hop}=H^{x}\cdot H^{x}+H^{y}\cdot H^{y}.
\end{equation}
Furthermore, $H^{\rm hop}\cdot H^{\rm hop}$, $H^{x}\cdot H^{x}$ and $H^{y}\cdot H^{y}$ all commute with each other and therefore they have common eigenfunctions. The eigenenergy ($E$) of $H^{\rm hop}$ can then be obtained from the simple Minkowski sum~\cite{zhu2020distinguishing},
\begin{equation}\label{eq8}
  E^2=E_x^2+E_y^2,
\end{equation}
with $E_x$ and $E_y$ being the respective eigenenergy of $H^x$ and $H^y$, sharing the same feature as the eigenenergy of $H^{\mathrm{SSH}}$ specified in Eq.~(\ref{eq4}).  Importantly, we have forced $H^{\mathrm{SSH}}$ in our design to have an odd number of sites, and so both $H^{x}$ and $H^{y}$ can only have one state with zero eigenvalue as the topological zero mode, with all others being bulk states with nonzero eigenenergy. Because the eigenenergy $E$ takes zero only when both $E_x$ and $E_y$ are zero, $H^{\rm hop}$ thus is constructed to only support one TCM with zero energy as shown in
Fig.~\ref{fig:delocalized}(a).  In addition, given that $H^{\rm hop}\cdot H^{\rm hop}$, $H^{x}\cdot H^{x}$ and $H^{y}\cdot H^{y}$ commute with each other, the obtained TCM can be shown to be a product state of the topological edge states of  $H^{x}$ and $H^{y}$, essentially two SSH models along with two perpendicular directions.   More details presented in Supplementary Material confirm that our model represents a TQI phase in terms of a  symmetry analysis  and the associated topological invariants.   Therefore, we now have an explicit expression of the TCM of a TQI phase, namely,
\begin{equation}\label{eq9}
  |\mathrm{corner},\mathrm{TQI}\rangle=|\mathrm{edge},\mathrm{SSH},x\rangle\otimes|\mathrm{edge},\mathrm{SSH},y\rangle.
\end{equation}
This explicit expression of a single TCM makes it possible to predict when the TCM will be completely delocalized in both directions, as shown below.

In a 1D SSH setting, a delocalized topological edge mode can be obtained by balancing the topological localization of the edge mode with NHSE~\cite{zhu2021delocalization}.  The latter should arise because there is already nonreciprocal hopping in our non-Hermitian system.
Because of our design, the $x$ and $y$ directions play an equal role in the TCM (see Eq.~\ref{eq9}), and therefore a delocalized TCM is also expected
when we choose a critical value of the non-reciprocity strength parameter $g$ for $H^{\rm hop}$.
The TCM thus obtained, though it tends to be localized at the top right corner of the square lattice, can now be made to be delocalized throughout all the ``BB" sites. One such example with $N=3$ is shown in Fig.~\ref{fig:delocalized}(b), with parameters set as $N=3$, $t_1=1$, $t_2=e$, $g=1$, $\gamma=0.5$, and $\eta=1$ (this set of parameters will be also used later as an example). Consistent with the fact that this TCM is built on 1D SSH zero edge modes which only occupy one sublattice due to the chiral symmetry of the SSH system, the TCM only occupies ``BB" sites even when it is delocalized, with the occupation density profile almost homogeneous across the entire 2D lattice.

{\it 2D topological laser array}---To uncover the consequences of our design, we first investigate the spectrum of the system in the linear limit $\eta\xrightarrow{}\infty$. The spectrum for different $\xi$ values is shown in Fig.~\ref{fig:delocalized}(a). As the value of $\xi$  increases, $\mathrm{Im}[E]$ of the TCM  changes from negative to positive. This clearly indicates that the lasing threshold for the TCM is around $\xi_{\rm thres}\approx1.3$.
Note also that the bulk spectrum changes slowly when increasing the pump strength $\xi$, thus promising single-mode lasing with a wider range of $\xi$.

A time-domain analysis is necessary to look into the stability of the lasing mode.  Consider then a nonlinear case with $\eta=1$. The dynamics of the system is governed by the Schrodinger's equation with an effective Hamiltonian, namely, $i\frac{d|\psi\rangle}{dt} = H|\psi\rangle$,
where $H$ is defined in Eq.~(\ref{eq1}), $|\psi\rangle$ denotes the state of the system, which can be expressed as a superposition $|\psi\rangle=\sum_{n} \psi_{n} |n\rangle$ with $\psi_n$ the state amplitudes.   We next solve this time evolution equation associated with a random initial state $|\psi^{\rm ini}\rangle=\sum_{n}\psi_{n}^{\rm ini}|n\rangle$, where $\psi_n^{\rm ini}$ is a random complex number.  After a sufficiently long period, the system evolves to a steady state, expected to be in the single-mode region~\cite{zhu2021single}.
We first check the total intensity $I_{\rm tot}=\sum_{n}|\psi_{n}|^2$ at the steady state, with the results shown in Fig.~\ref{fig:delocalized}(c). It is seen that the total intensity is zero when $\xi$ is smaller than the previously identified threshold and then linearly increases with $\xi$ after passing the threshold. The field density profile at the steady state should also be examined.  One computational example of the density profile with $\xi=2$ at the steady state is featured in Fig.~\ref{fig:delocalized}(d).  One observes that the density profile of the state field is indeed homogeneously distributed on ``BB" sites, just as the eigenfield shown in Fig.~\ref{fig:delocalized}(b).  This hence confirms that we can indeed obtain a lase array over a 2D geometry by making the TCM as the single lasing mode.   Returning to Fig.~\ref{fig:delocalized}(c), it is also seen that
for $\xi$ going beyond the lasing threshold, the total intensity output without the NHSE still increases linearly, which is consistent with previous findings \cite{zhang2020low,han2020lasing,kim2020multipolar,zhong2021theory,zhu2021single}. However, with NHSE (blue dotted line),  the increase of the total intensity output is much steeper than the case without NHSE (black squares).
This clearly indicates the benefit of delocalizing the TCM as the lasing mode to obtain a higher output to input ratio.
Notably, such lasing is activated by applying the optical gain only at the corner.

\begin{figure}[h]
    \includegraphics[width=\linewidth]{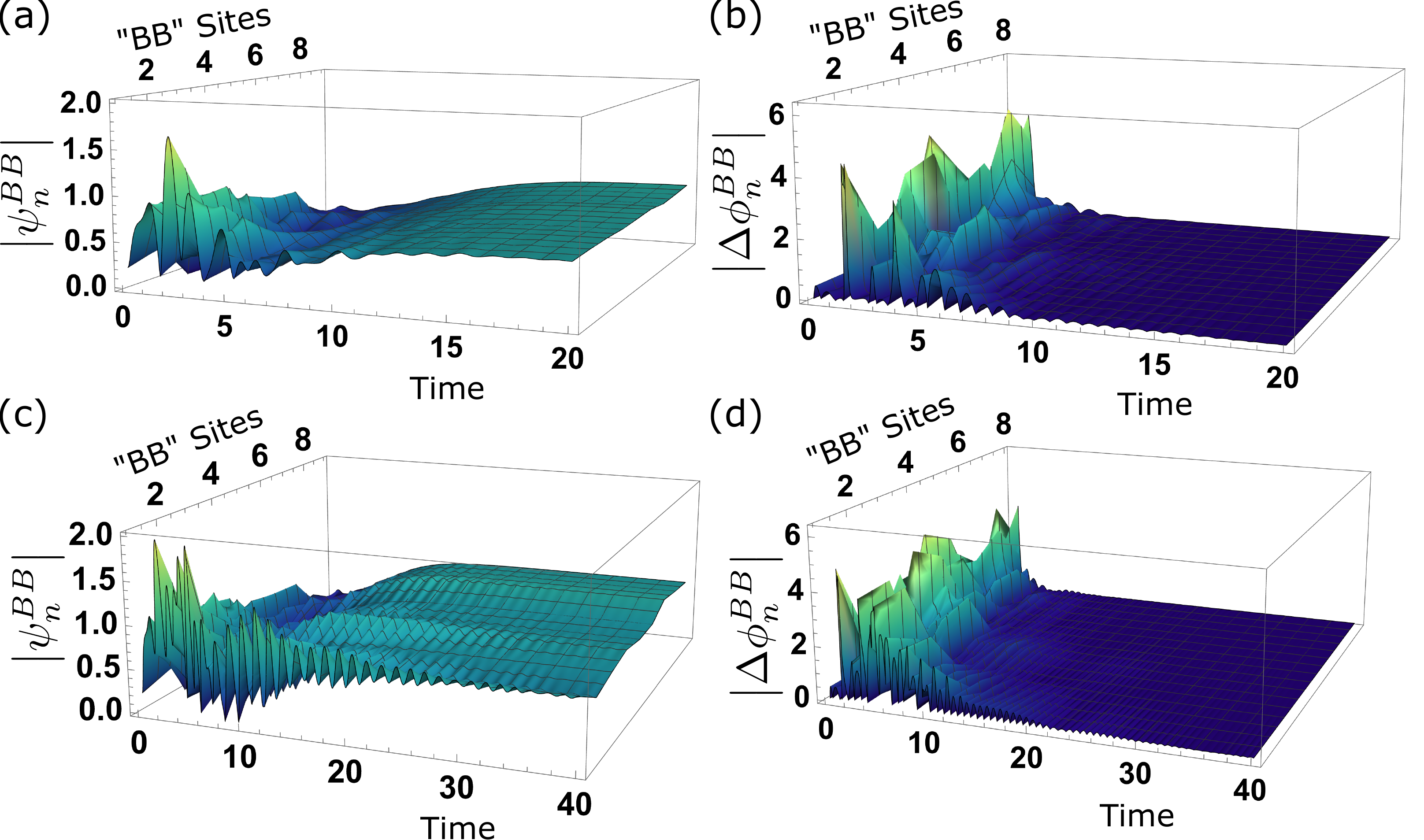}
    \caption[justification=justified]{(a) The time evolution of the magnitude of state amplitude over all the ``BB" sites, $|\psi_n^{BB}|$. They share almost the same amplitude after a long time. (b) The time evolution of the phase difference between two neighboring ``BB" sites, $|\Delta \phi_n^{BB}|$. The difference is always stabilized to 0. Here, we use an arbitrary initial state for the time evolution. (c)-(d) The same as (a)-(b), but we added weak disorder to the couplings of the system. The strength of disorder is 0.3 $(W=0.3)$.  Although it takes longer for the state to become stabilized, the state is still extended throughout ``BB" sites, and the phase difference is maintained at 0.}
    \label{fig:bb}
\end{figure}

To digest in more detail how the steady lasing is reached,  we next show in Fig.~\ref{fig:bb}(a) the time evolution of the state amplitudes on ``BB" sites $|\psi_n^{BB}|$ for $\xi=2$.  As shown in Fig.~\ref{fig:bb}(a),  after an initial period of irregular evolution due to the use of a random initial state,  the magnitudes of state amplitudes $|\psi_n^{BB}|$ on different ``BB" sites not only become stabilized, but also uniform.   In particular,  we check whether the laser emission from the 2D array of ``BB" sites is indeed phase-locked.  This can be quantitatively examined through the phase difference between neighboring ``BB" sites.
Let the phase of the amplitude $\psi_n$ at a given site $n$ defined as $\phi_{n}=\arg{(\psi_{n})}$. Then the phase difference between neighboring ``BB" site is then $\Delta \phi^{BB}_{n} = \phi^{BB}_{n+1} - \phi^{BB}_{n}$.  Our results on  $\Delta \phi^{BB}_{n}$ is presented in Fig.~\ref{fig:bb}(b).  Evidently, the initial phase differences are random, but all the state amplitudes are locked with zero phase difference when reaching the steady state. This is a highly desirable feature because the zero-phase-locked lasing can enhance the output intensity at the focus of the laser beams by a factor proportional to the square of the number of lasing elements in the laser arrays~\cite{Teimourpour2016}.
%A single-mode zero-phase-locked laser array is thus entire;y

The main advantage of using the TCM is the inherent robustness of our topological laser array.  To demonstrate this, we  add a disordered term $\delta$ to each coupling term $t_1$ and $t_2e^{\pm g}$, $\delta$ being a random number in the interval $[-0.5W,0.5W]$, and recheck the time evolution of the magnitude of the state amplitudes and the phase-locking behavior.
As shown in Fig.~\ref{fig:bb}(c) and (d), one can still obtain rather similar results, although the time required for the state to be stabilized is longer.
More discussions on disorder effects can be found in Supplementary Material.

\begin{figure}[h]
    \centering
    \includegraphics[width=0.9\linewidth]{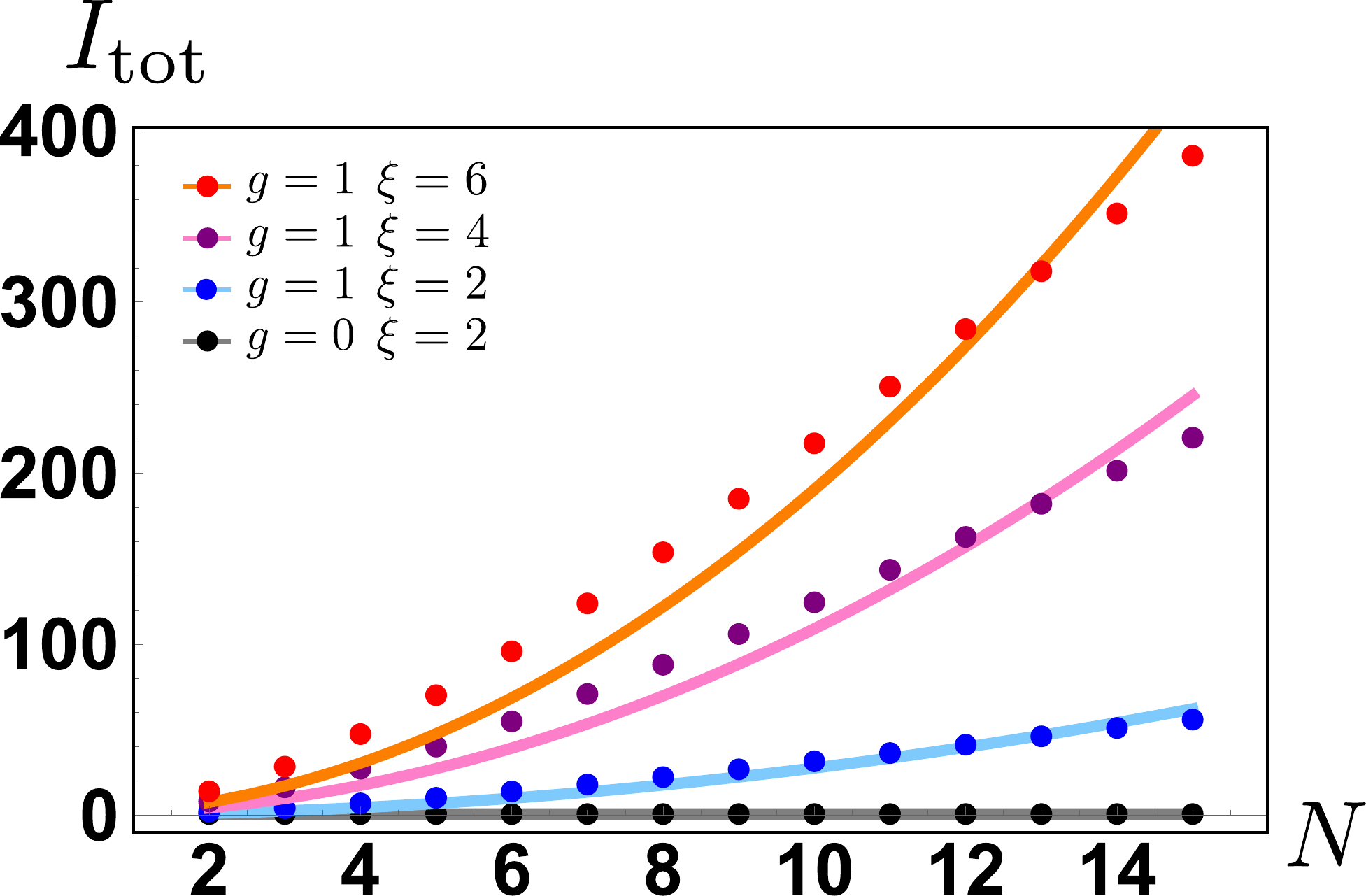}
    \caption{The total saturated intensity $I_{\rm tot}$ vs $N$ with different cases. The dots show the data obtained from numerical simulations, and the lines represent a fitting function, with the orange line being $I_{\rm tot}=1.90N^2$, the pink line being $I_{\rm tot}=1.09N^2$, the light blue line being $I_{\rm tot}=0.276N^2$, and the grey line being $I_{\rm tot}=0.693$. The power of the laser array is seen to scale with the area of the system with the assistance of NHSE. Besides, the power is also enhanced linearly by increasing the optical gain $\xi$.}
    \label{fig:com}
\end{figure}

In our calculations so far, we only considered $N=3$. However, one main motivation underlying our approach is to scale up the 2D phase-locked array. To that end, it is necessary to ensure that the working mechanism we propose has no issue for larger and larger 2D lattices.
This is verified by results with larger system sizes such as those shown in Supplementary Material.
Besides, we investigate here the optical intensity output in terms of a varying $N$, as shown in Fig.~\ref{fig:com} for cases with NHSE and a case without NHSE.  It is clear from Fig.~\ref{fig:com}  that with the assistance of NHSE, the total output intensity summed over all lattice sides indeed scales as $N^2$.
In addition, from the computationally calculated fitting curves, we can further deduce that the coefficient is linearly dependent on the optical gain, which is found to be $0.406(\xi-1.319)$.
This result coincides with what we have discussed previously in Fig.~\ref{fig:delocalized}(c), with the lasing threshold at $\xi_{\rm thres}=1.319$.
Therefore, after combining the effect of $\xi$ and $N$, we can obtain the following fitting relation depicting the output power: $I_{\rm tot}=0.406(\xi-\xi_{\rm thres})N^2$ for the set of parameters chosen here.
On the other hand, if there is no asymmetric coupling and hence no NHSE, the total output intensity remains constant even if the system size increases significantly.  These results strengthen our theoretical analysis above.

{\it Experimental consideration.}---Coupled optical ring resonators have already been investigated to realize both higher-order topological phases and NHSE. As such, this platform is suitable for the demonstration of a large-scale phase-locked 2D laser array. As shown above, the negative couplings in the system construction are essential to realize the TQI phase. Fortunately, this can be done by shifting the link rings~\cite{Mittal2019}. Further, the nonreciprocal couplings are also crucial to induce NHSE. Again, this aspect presents no issue either, by engineering gain/loss distribution on link rings~\cite{doi:10.1126/science.1258479,doi:10.1126/science.1258480,Zhu2020,Xiao2020,Weidemann2020}. The required loss can be realized by depositing Cr on the ring, while the gain can be realized by gain materials like InGaAsP~\cite{doi:10.1126/science.1258479,zhao2018topological,Peng2014}. Most reassuring is that the platform of coupled optical ring resonators has already been used to realize topological lasers in different studies~\cite{doi:10.1126/science.aar4005,zhao2018topological,zeng2020electrically,Choi2021,Smirnova2020}, though having not been used to realize a robust 2D laser array yet. Given that both optical pumped and electrically pumped topological lasers have been realized~\cite{Price2022},  all the important elements needed in our proposal are already available in existing experiments.

{\it Conclusion.}---
In conclusion, we have proposed a promising, experimentally feasible approach to the realization of single-mode zero-phase locked laser arrays that can be easily scaled up in two dimensions, with notable robustness.   The parameter regime for single-mode zero-phase-locked laser arrays is quite large and hence offers wide tunability. The output power is found to linearly increase with the single-site pump strength and also scales proportionally with the area of a coupled ring resonator setting.  

\begin{acknowledgements}
 {\bf Acknowledgements:} J.G. would like to acknowledge fund support by the Singapore Ministry of Education Academic Research Fund Tier-3 grant No. MOE2017-T3-1-001 and by the Singapore NRF Grant No. NRF-NRFI2017-04.
\end{acknowledgements}

\appendix

\section{Supplementary Material}
\subsection{Symmetry Analysis and Topological Invariants for Non-Hermitian Topological Quadrupole Insulator}
Here we carry out a symmetry analysis of the Bloch Hamiltonian of $H^{\rm hop}$ considered in the main text, together with its topological characterization.   Consider then the following Bloch Hamiltonian,
\begin{equation}
\begin{split}
  H(k_x,k_y)=&(t_1+t_2 e^g e^{-ik_x})\sigma_3\sigma_+ + (t_1+t_2 e^{-g} e^{ik_x})\sigma_3\sigma_-\\
  +&(t_1+t_2 e^g e^{-ik_y})\sigma_+\sigma_0 + (t_1+t_2 e^{-g} e^{ik_y})\sigma_-\sigma_0
  \end{split}
\end{equation}
where $\sigma_\pm=(\sigma_1\pm i\sigma_2)/2$. The system supports two pseudo mirror symmetries about $x$ and $y$ axes,
\begin{equation}
\begin{split}
    m_xH(k_x,k_y)m_x&=H^{\dagger}(-k_x,k_y),\\
     m_yH(k_x,k_y)m_y&=H^{\dagger}(k_x,-k_y).
\end{split}
\end{equation}
where $m_x=\sigma_0\sigma_1$ and $m_y=\sigma_1\sigma_3$. Note that $m_x$ anti-commutes with $m_y$, $\{m_x,m_y\}=0$.  This feature makes the energy bands of the system double degenerate~\cite{Benalcazar2017}.

\begin{figure}
	\includegraphics[width=\linewidth]{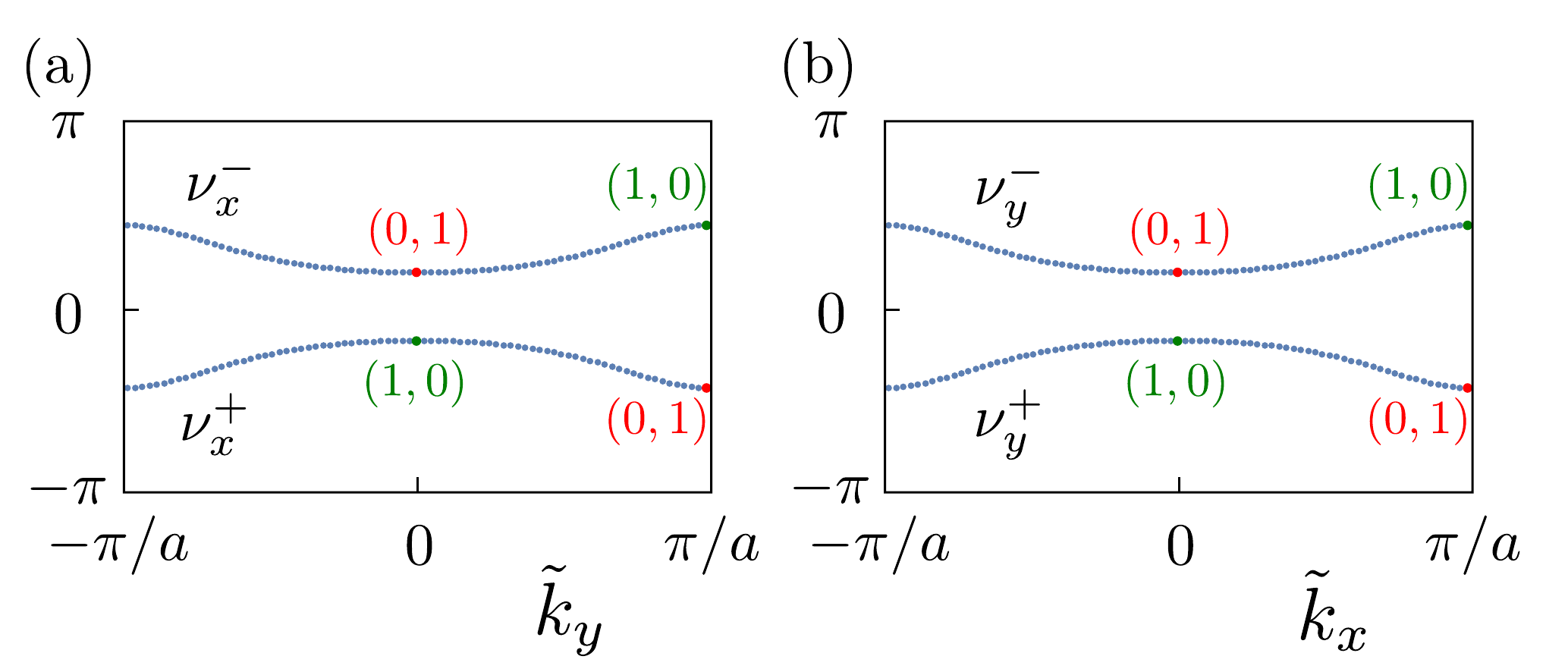}
    \caption[justification=justified]{ The Wannier bands for $H^{\rm hop}$ with $t_1=1$ and $t_2=e$. (a)(b) are the $x-$direction and $y-$direction Wannier bands as functions of $\tilde{k}_y$ and $\tilde{k}_x$.}
    \label{TQI}
\end{figure}

Because NHSE is present in our system, the usual bulk-boundary correspondence is broken. As such, the topological states are described by topological invariants defined on the so-called generalized Brillouin zone instead of the usual Brillouin zone [For our system, the generalized Brillouin zone can be obtained by $(\tilde{k}_x,\tilde{k}_y)=(k_x,k_y)+i(g,g)$]. For topological quadrupole insulators, the topological invariants are nested polarizations defined on the Wannier bands, which are the eigenvalues of Wilson-loop~\cite{Benalcazar2017,zhu2020distinguishing}. The Wilson loop is defined as follows,
\begin{equation}
  W_x(\tilde{k}_y)=\mathcal{P}e^{i\oint A^x_{mn}(\tilde{k}_x,\tilde{k}_y)d\tilde{k}_x},
\end{equation}
where $\mathcal{P}$ is the path ordering operator, $A^x_{mn}(\tilde{k}_x,\tilde{k}_y)=i\langle u_m(\tilde{k}_x,\tilde{k}_y)|\partial_{\tilde{k}_x}|u_n(\tilde{k}_x,\tilde{k}_y)\rangle$ is the non-Abelian Berry phase. For our system $ W_x(\tilde{k}_y)$ is a two-by-two matrix with eigenvalues $e^{i2\pi\nu_x^\pm}$. In Fig.~\ref{TQI}, we present the Wannier bands for our system.  It is seen that the two Wannier bands are symmetric around zero. The sum of Wannier bands corresponds to the polarization of the system so that the polarization of the system is zero. TQI is a higher-order topological insulator with zero polarization and quantized quadrupole moment. The quadrupole moment is determined by the topology of the Wannier bands, so-called nested polarization. By analyzing the eigenvectors at high symmetric momentum points of the Wannier bands, we observe that there is a band inversion at the $\Gamma$ point~\cite{zhu2020distinguishing}.  This suggest that the nested polarization is nonzero $p_y^{\nu_-}=1/2$ and $p_x^{\nu_-}=1/2$. The quadrupole moment of the system is $q_{xy}=2p_x^{\nu_-}p_y^{\nu_-}=1/2$.  These results finally confirm that the system considered here is topologically nontrivial and supports in-gap topological corner modes.  Certainly, when considering the system under the open boundary condition, we chose an odd number of lattice sites along each dimension to ensure that there is only one corner mode.

\subsection{Larger System Size $N$}
\begin{figure}[h]
    \includegraphics[width=\linewidth]{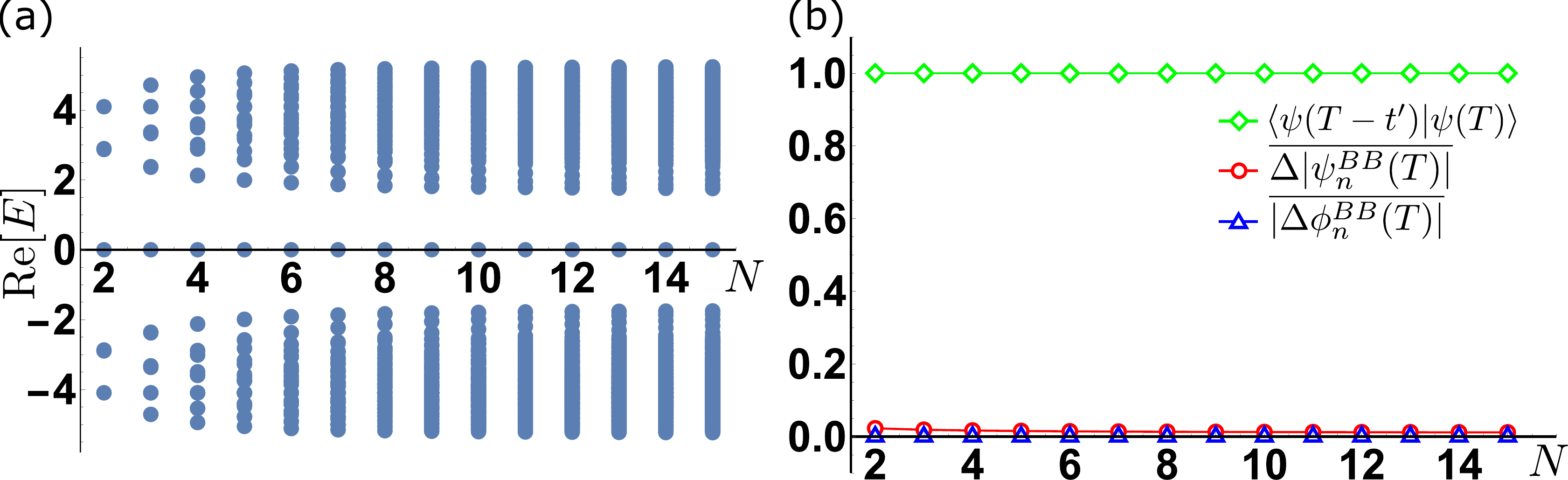}
    \caption[justification=justified]{(a) The real energy spectrum by increasing system size $N$. More bulk modes are observed when $N$ increases, but a single zero topological corner mode is always located within the band gap. (b) Several key properties vs system size $N$. They are the fidelity of the evolved states that obtained at two different evolution times, the mean of the difference in the magnitude of the state amplitudes between consecutive ``BB" sites, and the mean of the phase difference between consecutive ``BB" sites, which are shown as green diamonds, red circles, and blue triangles respectively. Their constant behaviors have shown that the evolved states of different system size $N$ are always stable, phase-locked, and extended throughout ``BB" sites after a long evolution time. System parameters here are set at $t_1=1$, $t_2=e$, $g=1$, $\gamma=0.5$, $\xi=2$, $\eta=1$, $T=200$, and $t'=4$.}
    \label{fig:ngen}
\end{figure}

In the main text, we have focused on the case of $N=3$.  Here we verify that the same mechanism works for  larger system sizes by tuning $N$.  First, we show the real part of the energy eigenvalues vs system size $N$ in Fig.~\ref{fig:ngen}(a).
As $N$ increases, the number of bulk modes increases accordingly, but there is always a single zero-mode located within the band gap, which is the topological corner mode.
Several key properties, such as the stability during long time evolution, phase-locking behavior, and the delocalization of the time-evolved state, are all verified by the results shown in Fig.~\ref{fig:ngen}(b).  That is, key system features remain unaffected when the system size $N$ increases.

\begin{figure}[h]
    \includegraphics[width=\linewidth]{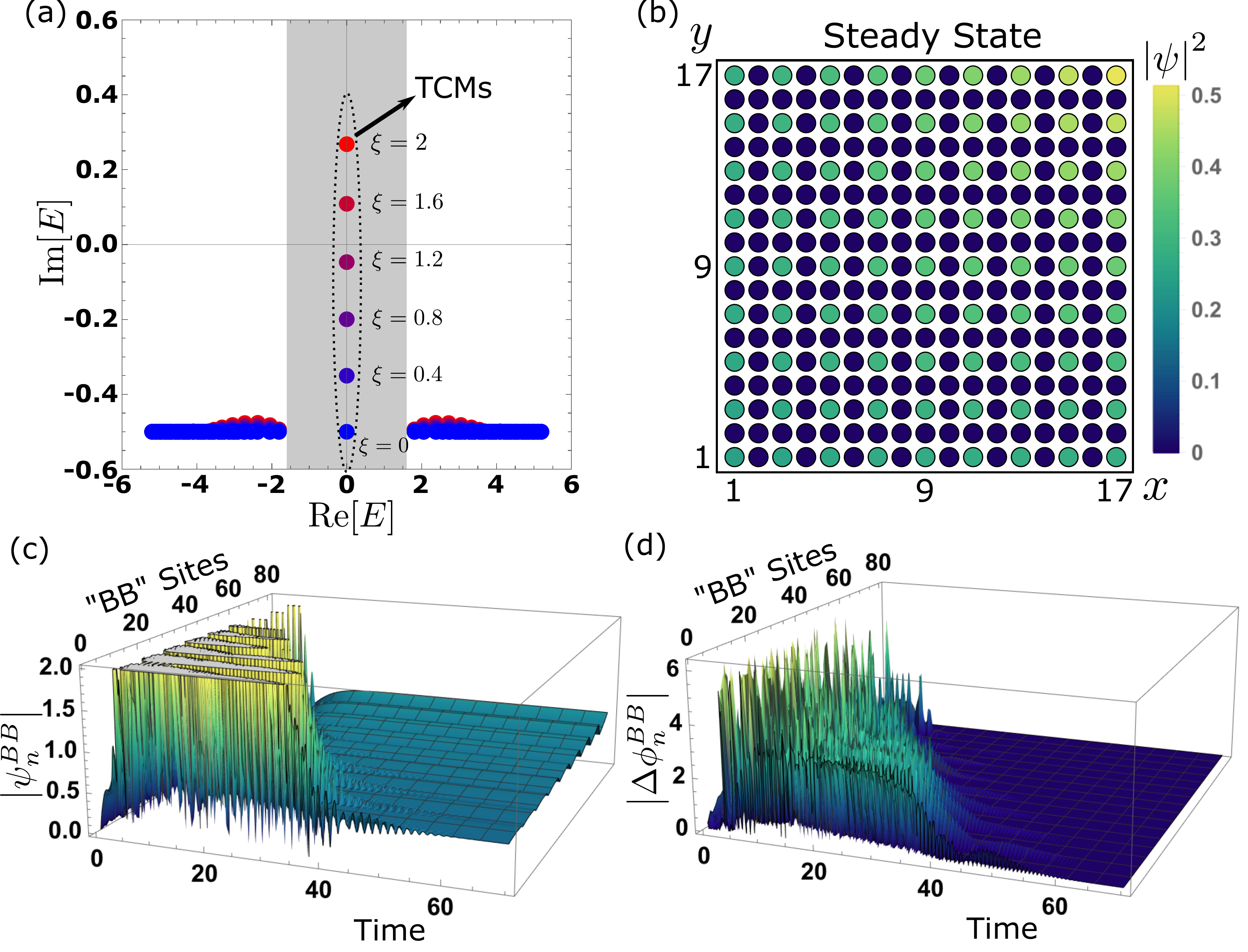}
    \caption[justification=justified]{Case study with $N=9$. (a) Energy spectrum with various values of the optical gain parameter $\xi$.  Results analogous to Fig.~\ref{fig:delocalized}(a) are observed, with more bulk modes. (b) The density profile of the steady state after a sufficiently long evolution time with $\xi=2$. The densities of the state are  distributed homogeneously over different ``BB" sites. (c) Time evolution of the magnitudes of the state amplitudes over all the ``BB" sites, $|\psi_n^{BB}|$. They are stabilized and of the same value after a long time. (d) Time evolution of the phase differences between neighboring ``BB" sites, $|\Delta \phi_n^{BB}|$. The difference is stabilized to 0 after a long time. Unless otherwise specified, the chosen system parameters are $N=9$, $t_1=1$, $t_2=e$, $g=1$, $\gamma=0.5$, $\xi=2$, and $\eta=1$.}
    \label{fig:n9}
\end{figure}

We also provide another concrete example with $N=9$ here to elaborate on some details.
In Fig.~\ref{fig:n9}(a), we show the energy spectrum with various values of the optical gain parameter $\xi$, showing features
analogous to the energy spectrum for the case of $N=3$ (Fig.~\ref{fig:delocalized}(a)), with more bulk modes.
The density profile of the steady state after sufficiently long time evolution with $\xi=2$ is also presented in Fig.~\ref{fig:n9}(b), showing that it is still extended throughout all the ``BB" sites.
We further investigate the time evolution of the magnitudes of the state amplitudes and the phase differences between two neighboring ``BB" sites in Fig.~\ref{fig:n9}(c) and (d), respectively.
A stable, uniformly distributed, and phase-locked steady state is also obtained.

In summary, our key results in the main text are also computationally observed in systems with larger $N$.  This is the reason why our proposed design for phase-locked laser arrays can be scaled up.

\subsection{Performance of 2D Laser Array with Other System Parameters}
\begin{figure}[h]
    \includegraphics[width=\linewidth]{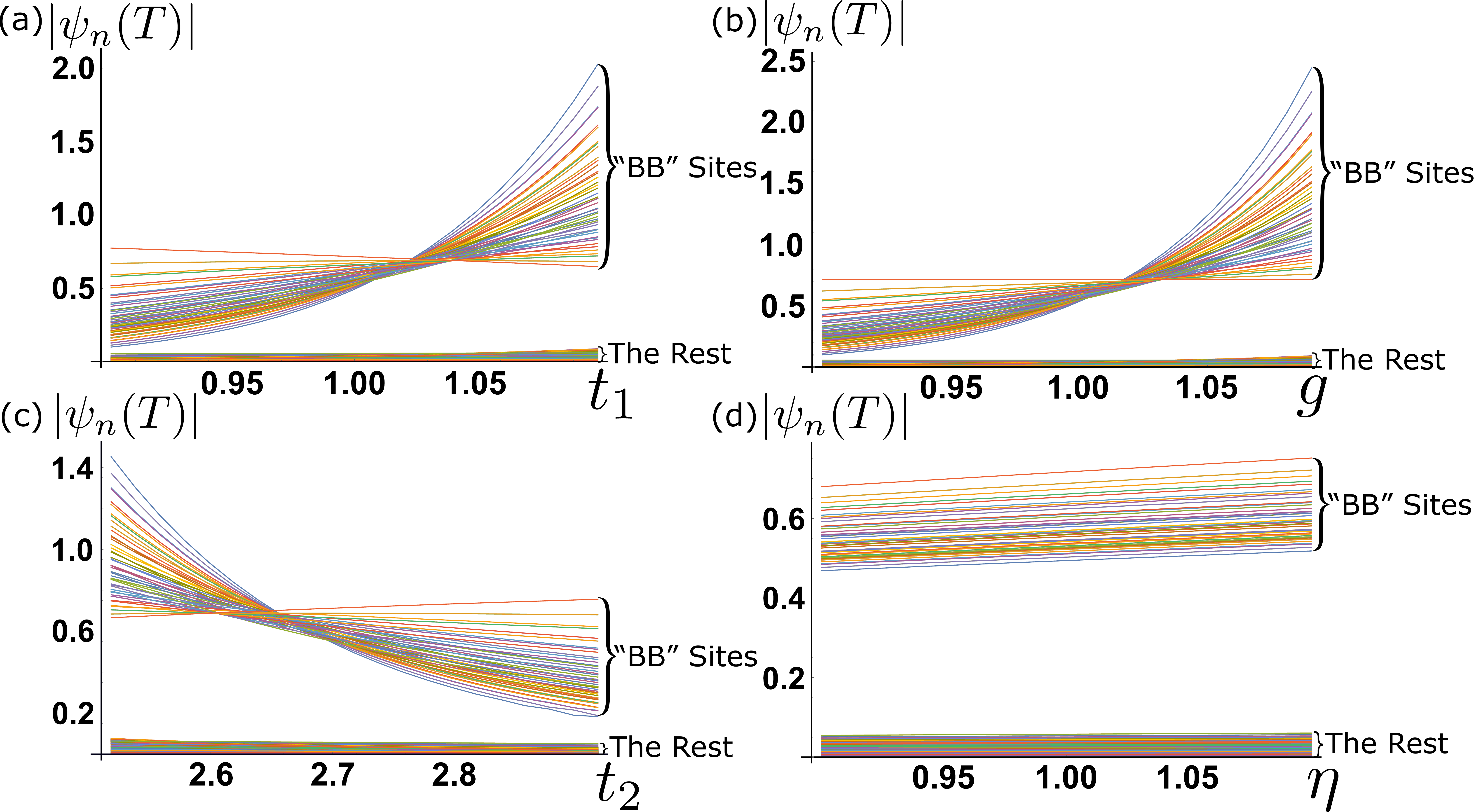}
    \caption[justification=justified]{Performance of 2D laser array with other system parameters, in terms of saturated magnitudes of the state amplitudes on each site, $|\psi_n(T)|$, with varying intracell couplings $t_1$ (a), the strength of the non-reciprocal hopping parameter $g$ (b), intercell couplings $t_2$ (c) and nonlinearity $\eta$ (d). Unless specified otherwise, parameters are chosen the same as Fig.~\ref{fig:n9}.}
    \label{fig4}
\end{figure}

In this section, we present some interesting results as we explore a wider range of system parameters. Figure~\ref{fig4} depicts how the saturated magnitudes of the state amplitudes, $|\psi_n(T)|$, are affected by varying system parameters.
The stimulated response is found to persist throughout the ``BB" sites even the important non-reciprocal parameter $g$ is tuned away from its most ideal values.
Further, as shown  Fig.~\ref{fig4}(a)-(c),  one can obtain a more extended response with larger $g$, $t_1$ or smaller $t_2$. In particular, due to the loss of the resonators considered, $|\psi_n(T)|$ in the bulk are slightly smaller than at the corner. Hence, by enhancing the strength of the skin effect or by reducing the localization strength of the TCM, we can overcome this  minor effect.
Moreover, in  Fig.~\ref{fig4} (d), we observed that by increasing $\eta$, the amplitude of the response could be amplified accordingly.

\subsection{Effect of Disorder}

It is necessary to investigate how disorder might affect the performance of the proposed 2D laser array.  In particular,  perfect modulation of the coupling strengths considered in our system construction is impossible.
We examine here the robustness of our results against coupling disorder.
To that end, we deliberately add a disordered term $\delta$ to each coupling term $t_1$ and $t_2e^{\pm g}$ in the system and randomly choose $\delta$ from a random number in the interval $[-0.5W,0.5W]$, where $W$ represents the disorder strength.
In the presence of disorder, we computationally solve the time evolution of a given initial state, and denote the evolved state as $|\psi_d(T)\rangle$.

\begin{figure}
	\includegraphics[width=\linewidth]{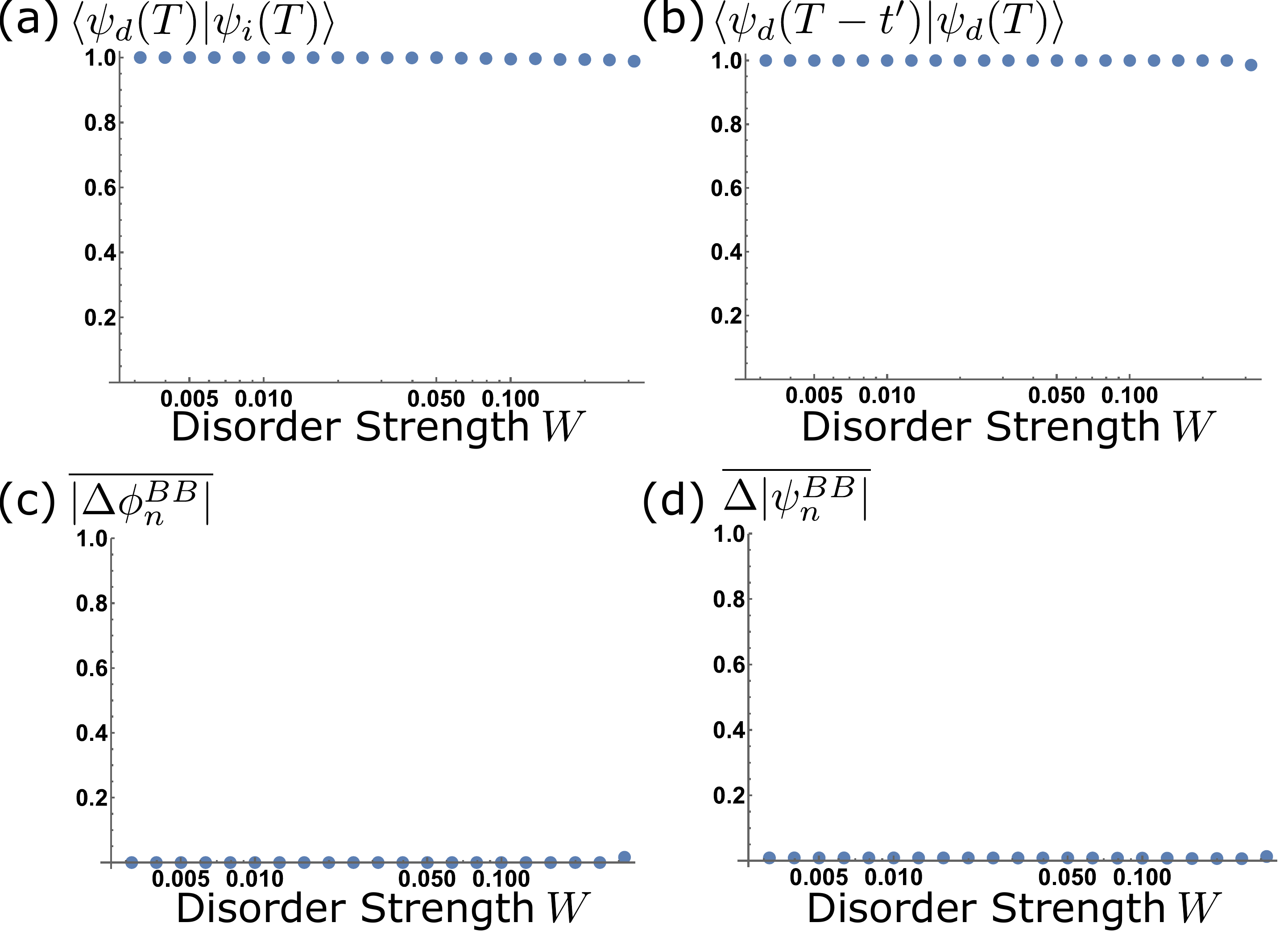}
    \caption[justification=justified]{ (a) The fidelity between the evolved state with and without the disorder. The fidelity is always very close to unity with small disorders introduced.  (b) The fidelity of the evolved states with the disorder, obtained under two different evolution histories over a long time.  This indicates the stability of the time evolution with the presence of disorder. (c) The mean of the phase difference of the evolved state between neighboring ``BB" sites with the disorder.  Results here verify that the evolved state with the disorder is phase-locked, in the presence of considerable disorder.  (d) The mean of the differences in the magnitude of the state amplitudes between neighboring ``BB" sites, for an evolved state with the disorder.  The results indicate that the evolved state is still delocalized over ``BB" sites.  All results here are obtained by averaging over 200 disorder realizations. Other system parameters are chosen as $N=3$, $t_1=1$, $t_2=e$, $g=1$, $\gamma=0.5$, $\xi=2$, $\eta=1$, $T=100$, and $t'=4$.}
    \label{fig:disorder}
\end{figure}

First, we compare the results with and without the disorder, and denote the latter as $|\psi_i(T)\rangle$.  Their overlap vs a varying disorder strength $W$ is shown in Fig.~\ref{fig:disorder}(a).  There it is seen that the fidelity is always close to unity, showing that the disorder does not affect much. Next, we check the stability of the evolved state with the disorder in Fig.~\ref{fig:disorder}(b), in terms of its fidelity with the evolved state at a different time, denoted $|\psi_d(T-t')\rangle$.
Furthermore,  we observe that the evolved state with the disorder is still phase-locked and delocalized over ``BB" sites.  We hence
evaluate the averaged phase difference between neighboring ``BB" sites and see that it is always close to $0$ when the disorder is accounted for, as shown in Fig.~\ref{fig:disorder}(c).
The averaged difference in field magnitudes between neighboring ``BB" sites also  remains at a small value, as shown in Fig~\ref{fig:disorder}(d).
Note also that  Fig.~\ref{fig:bb}(c)-(d) in the main text also presents one computational example illustrating the robustness to disorder.   With these results, one can safely conclude that the performance of the proposed 2D laser array is robust against weak disorder.

%\input{sections/acknowledgements.tex}

%\begin{thebibliography}{4}
%\bibliographystyle{unsrt}
%\bibliography{reference}
%\end{thebibliography}


\end{document}